\documentclass[authoryear,preprint,review,12pt]{elsarticle}

\usepackage{graphics}
\usepackage{graphicx}
\usepackage{epsfig}
\usepackage{amssymb}
\usepackage{subfigure}
\usepackage{rotating}
\usepackage{longtable}
\def\astrobj#1{#1}

\journal{New Astronomy}

\begin{document}

\begin{frontmatter}





\title{The Absolute Parameters of The Detached Eclipsing Binary V482 Per}


\author[label1]{\"O. Ba\c{s}t\"urk\corref{cor}}
\ead{obasturk@ankara.edu.tr}
\author[label2,label3]{S. Zola}
\author[label4]{A. Liakos}
\author[label5]{R.~H. Nelson}
\author[label6]{K. Gazeas}
\author[label1]{\.{I}. \"{O}zavc{\i}}
\author[label1]{M. Y{\i}lmaz}
\author[label1]{H.~V. \c{S}enavc{\i}}
\author[label3]{and B. Zakrzewski}

\cortext[cor]{Corresponding author. Tel.: +90 312 2126720; fax +90 312 2232395}

\address[label1]{Ankara University, Faculty of Science, Department of Astronomy and Space Sciences, TR-06100, Tando\u{g}an, Ankara, Turkey}
\address[label2]{Astronomical Observatory, Jagiellonian University, ul. Orla 171, PL-30-244 Krakow, Poland}
\address[label3]{Mt Suhora Observatory, Pedagogical University, ul. Podchorazych 2, PL-30-084 Krakow, Poland}
\address[label4]{Institute for Astronomy and Astrophysics, Space Applications and Remote Sensing, National Observatory of Athens, Penteli, Athens, Greece}
\address[label5]{1393 Garvin Street, Prince George, BC V2M 3Z1, Canada}
\address[label6]{Department of Astrophysics, Astronomy and Mechanics, National \& Kapodistrian University of Athens, Zografos, Athens, Greece}

\begin{abstract}

We present the results of the spectroscopic, photometric and orbital period 
variation analyses of the detached eclipsing binary \astrobj{V482~Per}. We derived 
the absolute parameters of the system (M$_{1}$ = 1.51 M$_{\odot}$, 
M$_{2}$ = 1.29 M$_{\odot}$, R$_{1}$ = 2.39 R$_{\odot}$, R$_{2}$ = 1.45 R$_{\odot}$, L$_{1}$ = 10.15 L$_{\odot}$, L$_{2}$ = 3.01 L$_{\odot}$) for the first time
in literature, 
based on an analysis of our own photometric and spectroscopic 
observations. We confirm the nature of the variations observed in the 
system's orbital period, suggested to be periodic by earlier works. 
A light time effect due to a physically bound, star-sized  companion 
(M$_{3}$ = 2.14 M$_{\odot}$) on a highly eccentric (e = 0.83) orbit, seems 
to be the most likely cause. We argue that the companion can not be a
single star but another binary instead.  
We calculated the evolutionary states of the system's components, and
we found that the primary is slightly evolving after the  Main Sequence, while
the less massive secondary lies well inside it.
\end{abstract}


\end{frontmatter}


\section{Introduction}
\label{}

The mass of a star is the single most important parameter which determines the 
way it lives and dies and what remains from it after its death. It can only be 
directly and precisely determined with the analysis of the light and radial 
velocity variations occurring due to the orbital motion of the components of 
an eclipsing binary system. Since each component of a binary 
is confined to the space that we call the Roche lobe, 
its evolution will be different 
from that of a single star once it fills this space. Mass transfer occurs 
between the two after contact has been made and disturbs 
the evolution of the components 
as single stars. Detached binaries are the systems whose components have not 
filled their Roche lobes yet and thus evolve as single stars. Therefore the 
components of detached eclipsing binaries with well determined 
parameters have been regularly used to test the stellar evolution models.  

\astrobj{V482 Per} (BD+47$^{\circ}$ 961, GSC 3332-314) is a detached eclipsing 
binary which was first observed and classified as a short period variable by 
\citet{hoff66}. \citet{harvig81} investigated this star based on archival 
photographic plates. They classified it as an eclipsing system, found its magnitude, 
and published the first light-curve and its elements. \citet{agererlicht91}, based on 
their first photoelectric observations, published B and V light curves and the 
B-V color curve of \astrobj{V482 Per}.  \citet{popper96} found a mean spectral 
type of F2 for \astrobj{V482 Per} by using equivalent widths of the 
Na-D line of both components. 
In the same study he pointed to the discrepancy between the 
photometric study by \citet{agererlicht91}, which implied a middle-F spectral 
type for the primary, and the weakness of the Na-D lines in his spectra from 
which he determined an earlier spectral type. \astrobj{V482 Per} was 
also found to be a close binary in a triple system 
from the light-time effect observed 
on its O-C diagram. This subject was discussed  by \citet{wolfetal04} and 
\citet{ogloza12}, 
both of whom gave very similar values of the basic parameters 
of the third body ($P_{3}$ and e). 

In 2011, we acquired precise photometric measurements of the system 
in Gerostathopoulion Observatory of the University of Athens. We also obtained 
spectroscopic data for the system in order to determine its orbital parameters 
and the ratio of the masses of its components. We performed an analysis using the Wilson-Devinney 
\citep{wilsondev71,wilson90,kallrath98} model of our light curves 
based on our spectroscopically determined mass ratio of the system. 
As a result, for the first time we derived 
its absolute parameters (masses, radii, and luminosities) for both of 
the spectral types (F2 and F5) suggested by earlier works of \citet{agererlicht91} 
and \citet{popper96}. For the best fit we needed to include 
a third light, which would also support the triple-system 
hypothesis of earlier works by  
\citet{wolfetal04} and \citet{ogloza12}. We have performed an O-C analysis to 
ensure that an expected orbital variation is actually observed. 
Consequently, we confirmed the existence of a physically bound, 
stellar size third body and determined its parameters.

\section[]{Observations and Data Reduction}

Between October and December 2011 (in 13 nights), we obtained multicolor photometric 
data for \astrobj{V482~Per} with the 40 $cm$ Cassegrain telescope located at the 
Gerostathopoulion Observatory of the University of Athens. 
The telescope's focal ratio was converted to f/5.1 from f/8 by using a focal reducer. It was equipped with
an SBIG ST-10 XME CCD detector and a set of Johnson-Cousins $BVRI$ filters. 
We present a log of these observations in Table-\ref{table1}. 
We have corrected all the CCD images for instrumental effects,
using the C-munipack \citep{hroch98, motl04} software package. 
Next we performed differential apperture photometry with the same 
software, which allowed us to experiment with different aperture sizes 
until we achieved a relatively low scatter. 
Individual differential magnitudes were derived with respect to a chosen comparison 
star (\astrobj{GSC~3332-2173}), the stability of which was tested against a check star (\astrobj{GSC~3332-1993}).
We constructed the light curves using these differential magnitudes in each filter
and the light elements that we calculated from a linear fit to the previously 
published times of minima. 

\begin{table*}[t]
\scriptsize
\begin{center}
\caption{Information About The Observations}
\label{table1}
\medskip
\begin{tabular}{lcc}
\hline
Log of The Observations  & & \\
\hline
Dates of Observation (in 2011)  & \multicolumn{2}{c}{17,19,20,31 Oct}   \\
 & \multicolumn{2}{c}{1, 2, 3, 8, 16, 19, 26 Nov}  \\
 & \multicolumn{2}{c}{1, 4 Dec}  \\
\# of Data Points & \multicolumn{2}{c}{1630 (B), 1597 (V), 1614 (R), 1549 (I)} \\
Mean Errors ($\sigma$) & \multicolumn{2}{c}{0.0031 (B), 0.0031 (V), 0.0031 (R), 0.0027 (I)} \\
\hline
Star & m$_{V}$ (mag) & B-V (mag) \\
\hline
Variable (V482~Per) & 10.30 & 0.40\\
Comparison (GSC 3332-2173) & 9.63 & 0.32\\
Check (GSC 3332-1993) & 11.1 & 0.40\\
\hline
Light elements employed to phase the observations & & \\
\hline
Reference Epoch (in HJD) & \multicolumn{2}{c}{2455868.4005} \\
Orbital Period & \multicolumn{2}{c}{2$^{d}$.44677} \\
\hline
\end{tabular}
\end{center}
\end{table*}

\begin{figure}
\begin{center}
\includegraphics[width=11.2cm]{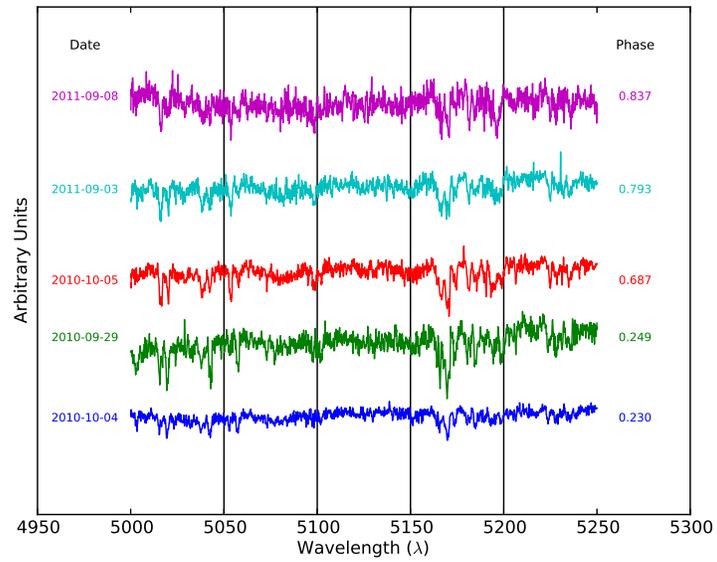}
\end{center}
\caption{The spectra used in radial velocity analysis. Four individual spectra 
are shifted arbitrarily for clarity.}
\label{fig1}
\end{figure}

One of the authors of this study (RHN), carried out spectroscopic observations
at the Dominion Astrophysical Observatory (DAO) in Victoria, British Columbia, 
Canada, using the Cassegrain spectrograph attached on the 1.85 m Plaskett telescope 
with a  grating (\#21181) of 1800 lines/mm, blazed at 5000 \AA~giving a reciprocal 
linear dispersion of 10 \AA/mm in the first order and covering a wavelength region 
from 5000 to 5260 \AA. The reductions (cosmic hit removal, median background 
fitting and subtraction for each wavelength, aperture summation, and continuum 
normalization) were performed with the 'RaVeRe' software \citep{nelson2010a} . 
The same software was used for wavelength calibration and linearization 
using the Fe-Ar spectra as wavelength standards. 
The five spectra we used are depicted in Fig.~\ref{fig1}

\begin{figure}
\begin{center}
\includegraphics[width=11.2cm]{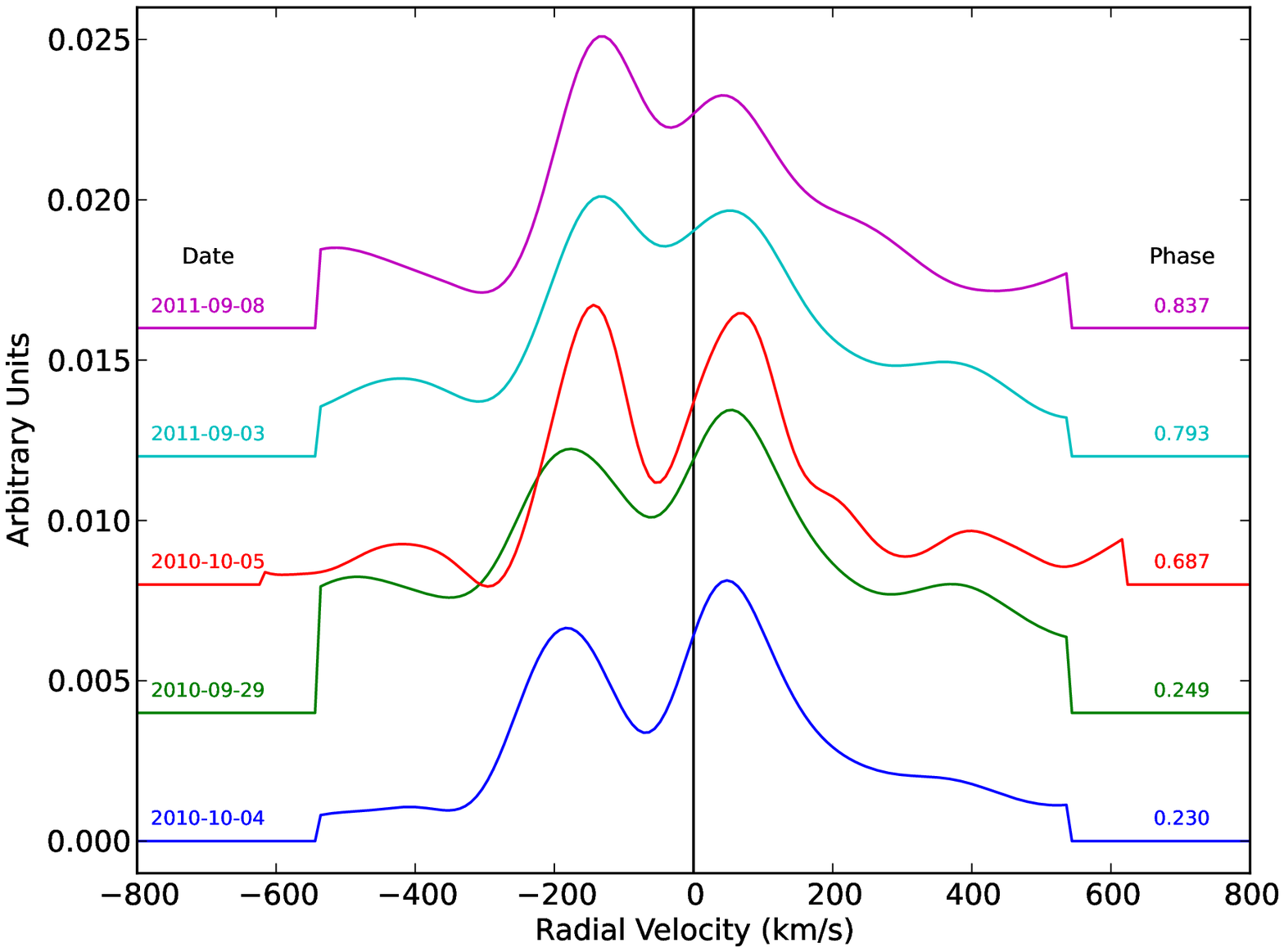}
\end{center}
\caption{Broadening functions used to determine radial velocities. 
Some of the functions are shifted arbitrarily for clarity.}
\label{fig2}
\end{figure}

\begin{figure}
\begin{center}
\includegraphics[width=11.2cm]{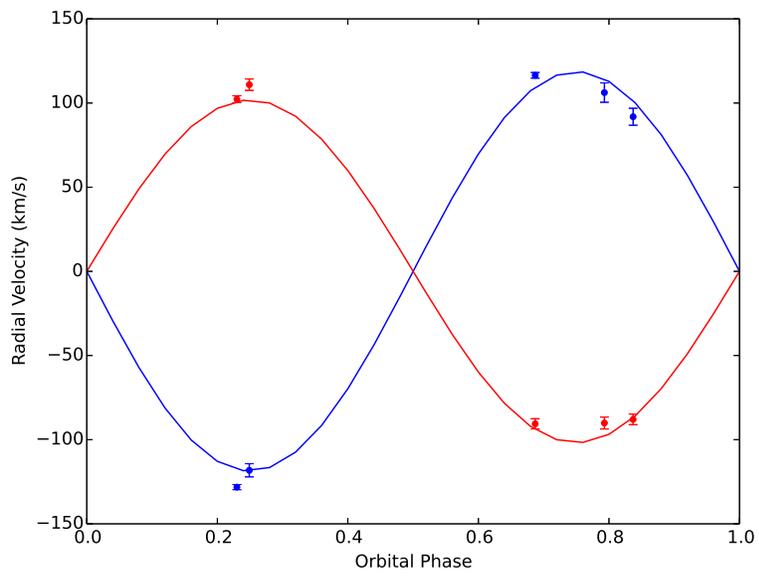}
\end{center}
\caption{Radial velocity data for \astrobj{V482~Per} and the best fits in solid lines.}
\label{fig3}
\end{figure}

In order to obtain radial velocity measurements we applied the Rucinski 
broadening functions \citep{rucinski04,nelson2006,nelson2010b} (Fig. \ref{fig2}). 
We phased these observations with the same light elements that we used to calculate
orbital phases of our photometric measurements, after having converted all the 
observation times to heliocentric values. We found the gamma velocity (V$_{\gamma}$) 
of the system to be -22.72 ($\pm$3.05) km/s. 
The resultant radial velocity curves for each of the components and the best fits to 
them are given in Fig. \ref{fig3}.
 We  determined semi-amplitudes of each of the 
fitted curves (K$_{1}$ and K$_{2}$) and computed the spectroscopic mass ratio 
of the system (q = m$_{s}$/m$_{p}$ = K$_{1}$/K$_{2}$ = 0.856).

\section{The Light Curve Analysis}
We used the Wilson-Devinney code (WD) \citep{wilsondev71,wilson90,kallrath98} 
to derive  parameters of components of \astrobj{V482~Per}. 
Instead of using all the individual observations, 
approximately 150 mean points  were used to speed up computations. 
They were calculated in such a way that 
they evenly covered the observed light curve. 
In order to make sure that the global minimum was found 
within the set ranges of free parameters, we made use of the Monte Carlo search 
algorithm. We followed the procedure outlined in \cite{zolaetal2014}, 
in which the mass ratio is fixed to the spectroscopically determined value.  
The albedo and gravity darkening coefficients were set at their theoretical 
values for a convective envelope (0.5 and 0.32, respectively) suitable for both
F2 and F5 spectral types \citep{vanHamme93}. 
The limb darkening coefficients were taken from the tables 
by \cite{Claret2011}, \cite{Claret2012} and \cite{Claret2013} and were built into the code. 
They 
were selected according to the temperature and the wavelength of the 
observations. We adjusted the inclination, temperature of the secondary, 
surface potentials, 
and the luminosity of the primary component. We made checks for all three existing
spectral types of this systems: F2 given by  \citet{popper96}, F5 derived
by \citet{agererlicht91} from the B-V color and A0 listed in the SIMBAD 
database.   
Due to a possibility that this system has a companion, third light was added
into the model. 
We fixed the temperature of the primary at the values of 6700~K,
6460~K and 9420~K, corresponding to the three spectral types according to 
the calibration given by \cite{Harmanec} and achieved three solutions. 
It soon turned out that the model with the assumed A0 spectral type of the primary
resulted in highly overluminous values for their masses. Therefore, this
solution was discarded. 
 
\begin{table*}
\scriptsize
\begin{center}
\caption{Best-fit parameters from Wilson-Devinney light curve modelling.}
\label{table2}
\medskip
\begin{tabular}{lcccc}
\hline
\hline
 & \multicolumn{2}{c}{F2} & \multicolumn{2}{c}{F5} \\
Stellar Parameters & Value  & Error & Value & Error  \\
\hline
$T_{1}$ [$K$] & 6700 & Fixed & 6460 & Fixed  \\
$T_{2}$ [$K$] & 6340 & 18 & 6130 & 17 \\
$\Omega_{1}$ & 5.38 & 0.04 & 5.39 & 0.04 \\
$\Omega_{2}$ & 7.42 & 0.15 & 7.38 & 0.15 \\
$q=m_{2}/m_{1}$ & 0.856 & Fixed & 0.856 & Fixed \\
$i \ [\circ] $ & 83.2 & 0.3 & 83.2 & 0.3 \\
\hline
Luminosities & & \\
\hline
$L_{1}$ [B] & 9.68 & 0.33 & 9.63 & 0.34 \\
$L_{1}$ [V] & 9.07 & 0.33 & 9.06 & 0.34 \\
$L_{1}$ [R] & 8.92 & 0.32 & 8.91 & 0.33 \\
$L_{1}$ [I] & 8.79 & 0.32 & 8.72 & 0.34 \\
$L_{2}$ [B] & 2.75 & 0.09 & 2.80 & 0.09 \\
$L_{2}$ [V] & 2.71 & 0.08 & 2.71 & 0.08 \\
$L_{2}$ [R] & 2.74 & 0.08 & 2.75 & 0.08 \\
$L_{2}$ [I] & 2.78 & 0.09 & 2.82 & 0.09 \\
$l_{3}$ [B] & 0.01 & 0.02 & 0.01 & 0.03 \\
$l_{3}$ [V] & 0.05 & 0.03 & 0.05 & 0.03 \\
$l_{3}$ [R] & 0.06 & 0.02 & 0.06 & 0.03 \\
$l_{3}$ [I] & 0.07 & 0.02 & 0.07 & 0.03 \\
\hline
\end{tabular}
\end{center}
\begin{flushleft}
\noindent {\bf Note:} $T_{1,2}$ - temperature of the primary and secondary,  
$\Omega_{1,2}$ - dimensionless surface potentials of the components, 
$q=m{\rm _2}/m{\rm _1}$ - the system mass ratio, $i$ - orbit inclination (in degrees),  
$L_{1,2}$ - WD code luminosities, $l_{3}$ - third light
\end{flushleft}
\end{table*}

The configuration of the system was found to be detached with both components
to be well within their Roche lobes.
We obtained a good fit to the observed light curves with just small discrepances
in the phase range between 0.30-0.45.
We derived a small contribution of the third light, ranging from 1 to 7\%,
depending on the passband.
 
\begin{figure}
\begin{center}
\includegraphics[width=12.8cm]{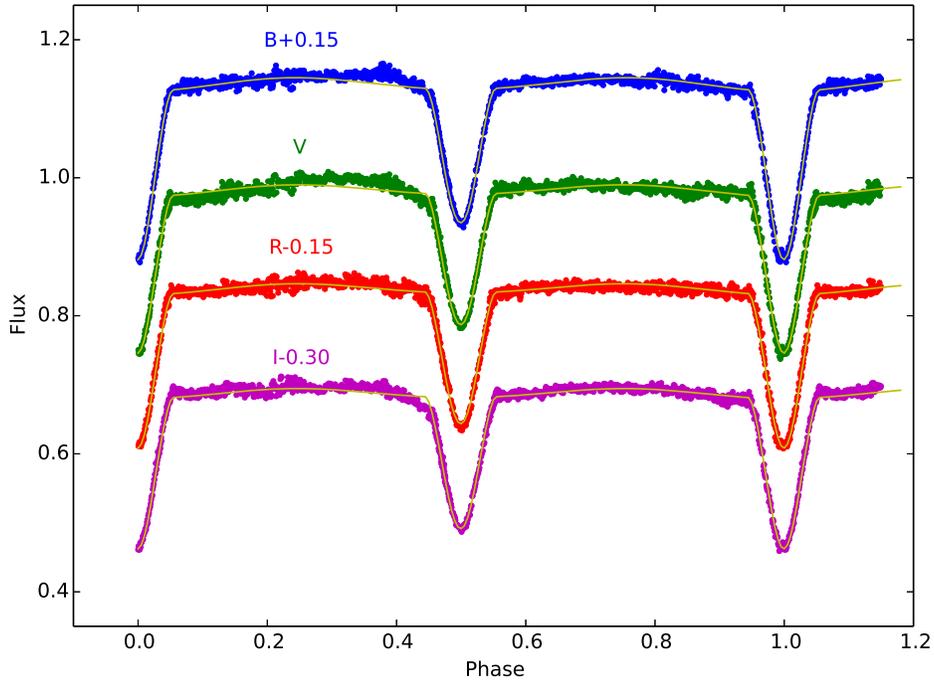}
\end{center}
\caption{Comparison between observations of  \astrobj{V482~Per} 
(BVRI filters - shown from top to bottom - shifted arbitrarily 
for clarity by the amounts given in the figure) and the best 
fits (solid lines) obtained for the F2 spectral type.
}
\label{fig4a}
\end{figure}

\begin{figure}
\begin{center}
\includegraphics[width=12.8cm]{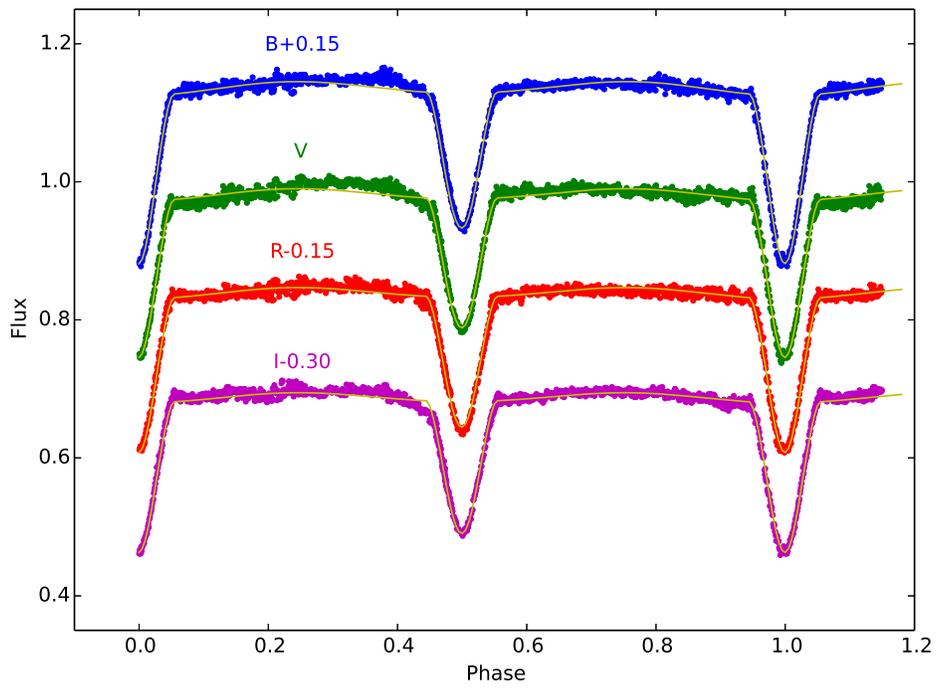}
\end{center}
\caption{The same as the Fig.\ref{fig4a} but for the spectral type F5. 
}
\label{fig4b}
\end{figure}

The results derived from the light curve modeling are presented in
Table \ref{table2}, while model light curves along with observations are shown in 
Figs. \ref{fig4a} \& \ref{fig4b} for both  F2 and F5 spectral types, respectively.

\section{Eclipse Timing Analysis}
\begin{figure}
\begin{center}
\includegraphics[width=14.2cm]{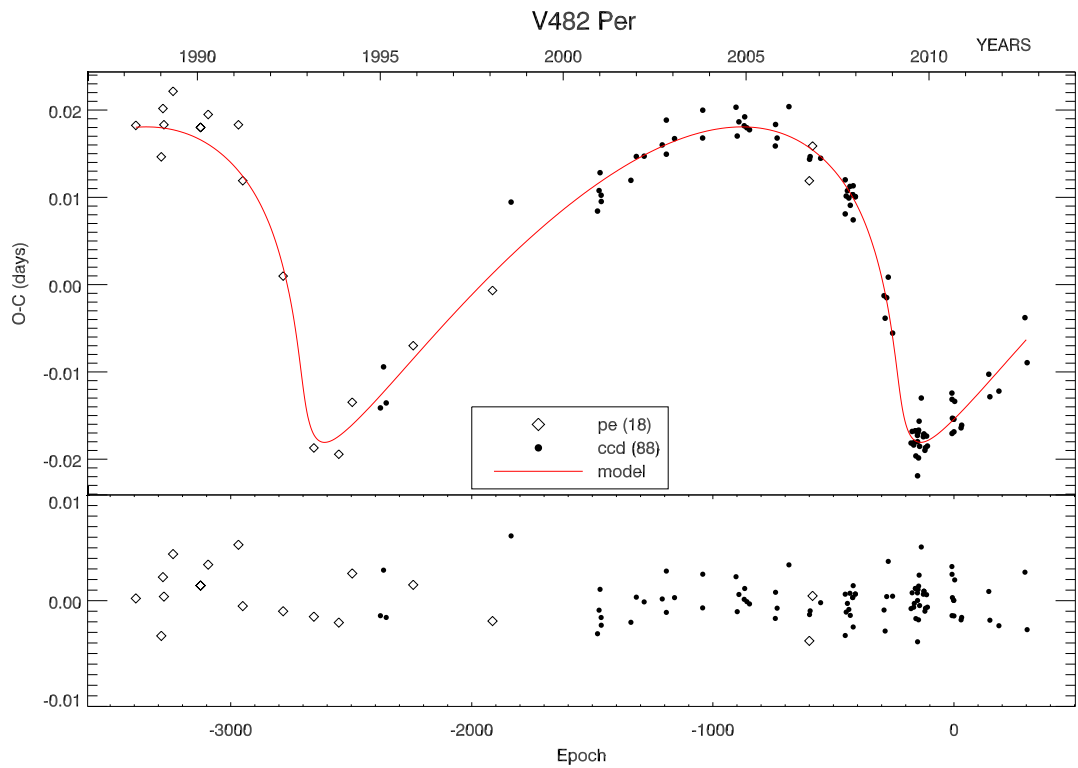}
\end{center}
\caption{Period variation behavior of \astrobj{V482~Per}. The best fit is shown as solid line.
Residuals from the fit are given in the lower panel. The symbols used have been defined in the legends. 
Times of minima are listed in the \ref{appendix} together with their references.}
\label{fig5}
\end{figure}

\astrobj{V482~Per} has been known to display orbital period
variations, which were 
suggested to be of a periodic nature \citep{wolfetal04, ogloza12}. We have 
collected all available eclipse times published in the literature (\ref{appendix}) 
including these determined from our data \citep{liakos2011} and constructed 
an O-C diagram (Fig. \ref{fig5}). 
We confirmed that the variations are in fact periodic and most likely they 
can be caused by the light time effect due to a third body gravitationally 
bound to the system on an eccentric orbit. We made use of Irwin's formalism \citep{irwin52,irwin59} during our analysis, and obtained the parameters of the third body presented in Table-\ref{table3} as a result.

\begin{table*}
\small
\begin{center}
\caption{Results of the eclipse timing analysis.}
\label{table3}
\medskip
\begin{tabular}{lcc}
\hline
\hline
Parameter & Value & Error \\ 
\hline
$P_{3}$ [years] & 16.56 & 0.11   \\
$A$ [days] & 0.018 &  0.001  \\
$\omega$ [$^{\circ}$] & 206 & 2  \\
$e$   & 0.83 & 0.02  \\
$a_{3}sin(i)$ [AU]   & 4.74 & 0.22  \\
$f(m_{3})$ [M$_{\odot}$]   & 0.39 & 0.06  \\
$m_{3}$ [M$_{\odot}$] & 2.14 & 0.05  \\
\hline
\end{tabular}
\end{center}
\end{table*}

\section{Results \& Discussion}
We have analyzed light and radial velocity curves, and the long term 
orbital period 
variations of the detached system \astrobj{V482~Per}. We have derived two sets 
of absolute parameters (Table-\ref{table4}) as a result of the first ever 
combined analysis of its photometric and spectroscopic observations.  
One set of solutions was based on the spectral type F5 given by 
\citet{agererlicht91}  and the other was based on F2 spectral type given by 
\citet{popper96}. We used the web interface to query Geneva Stellar Models' 
database \citep{ekstrom2012} 
and computed evolutionary tracks for three different masses (1.3, 1.5, 1.7 M$_{\odot}$) in close 
proximity to the derived masses of \astrobj{V482~Per}'s components 
(M$_{p}$ = 1.51 M$_{\odot}$, M$_{s}$ = 1.29). 
\begin{figure}
\begin{center}
\includegraphics[width=12.8cm]{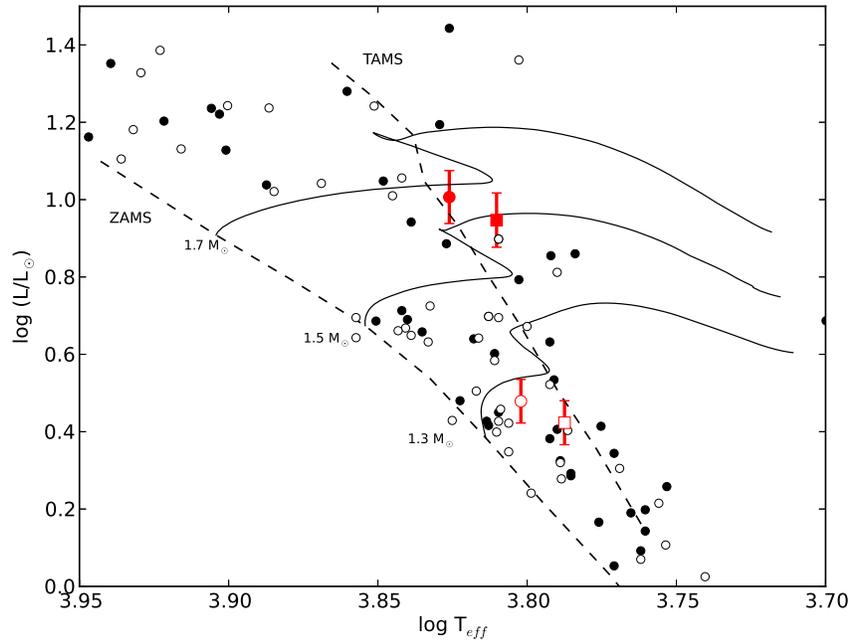}
\end{center}
\caption{A Hertzsprung-Russell Diagram with the positions of 
selected detached (primaries in filled, secondaries in 
unfilled grey circles) from \citet{torres2010}. Positions of the components 
of the system \astrobj{V482~Per} are marked with larger symbols (circles show 
F2-spectral type solution, squares that of F5). 
The evolutionary tracks for a dense grid of masses, 
computed by interpolation between the existing tracks of the Geneva stellar 
models \citep{mowlavi2012} are plotted with solid lines while ZAMS and TAMS 
are in dashed lines.}
\label{fig6}
\end{figure} 
The stellar evolutionary models have been taken from a dense grid for low mass 
stars assuming no rotation and for the solar metallicity \citep{mowlavi2012}.\\
We next constructed the Hertzsprung-Russell (HRD), 
Mass-Luminosity (MLD), and Mass-Radius (MRD) Diagrams (Figs. \ref{fig6} \& \ref{fig7}) 
based on the parameters we obtained from our analysis for \astrobj{V482~Per},
and the tracks from the models. We have given the positions of both 
components for each  of the two solutions in the HRD (Fig. \ref{fig6}) with filled (primaries) 
and unfilled (secondaries) circles (F2) and squares (F5).  The evolutionary model tracks 
for the three chosen masses are also drawn by solid lines in the HRD.
Since masses of the components do not depend on the assumed temperature of the 
primary, and both solutions resulted in almost the same values for the radii 
of components, only parameters resulting from the F2 spectral type model 
are presented in the MRD (right panel of Fig~.\ref{fig7}). 
However, due to the different temperatures of 
the components between the two solutions, there is a difference in 
luminosities of about 12\%, which is shown in the MLD by giving both 
of the solutions (left panel of Fig.~\ref{fig7}).

\begin{figure}[ht]
\centering
\begin{minipage}[b]{0.5\textwidth}
  \centering
  \includegraphics[width=75.00mm]{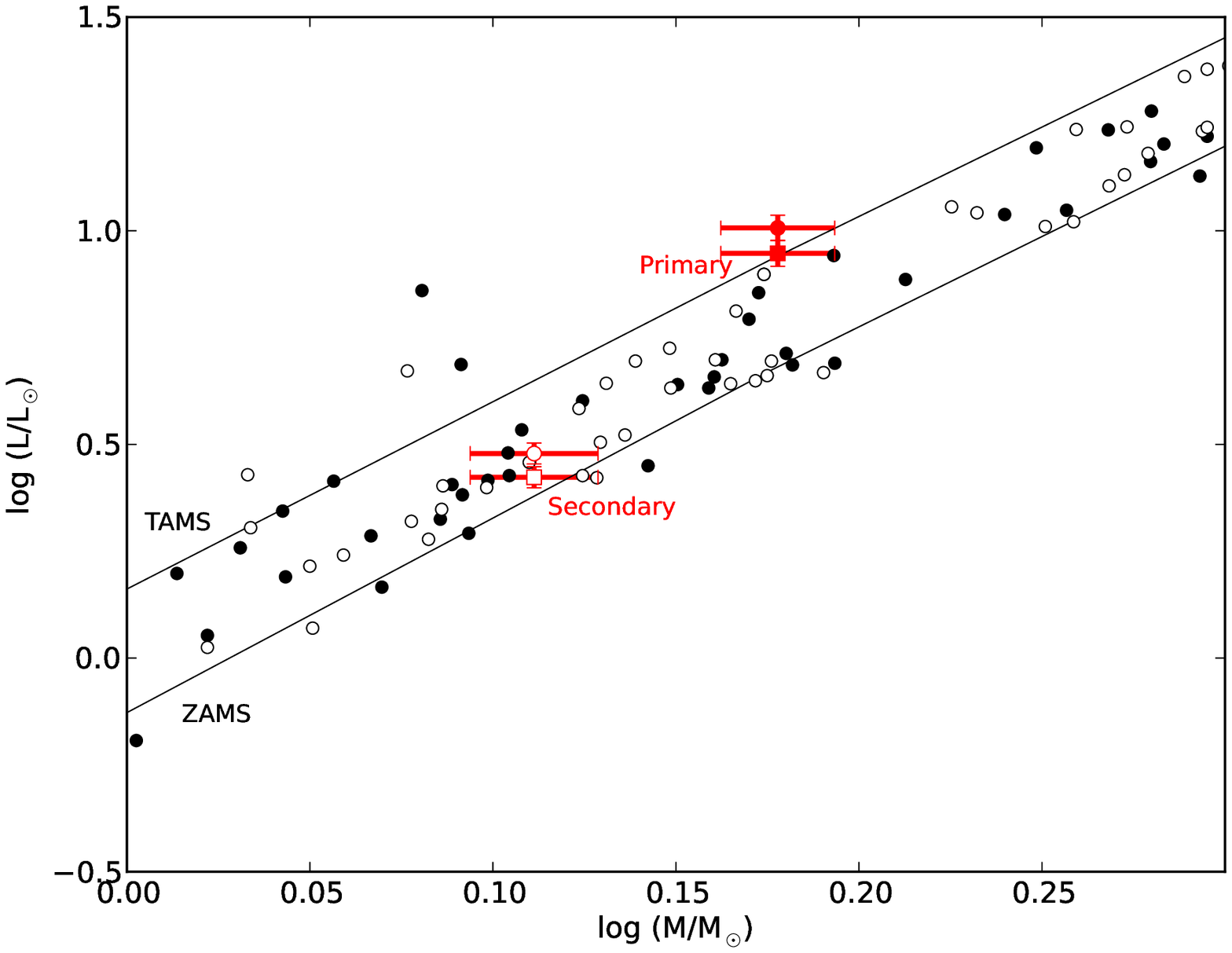}
  \label{fig7a}
\end{minipage}%
\begin{minipage}[b]{0.5\textwidth}
  \centering
  \includegraphics[width=75.00mm]{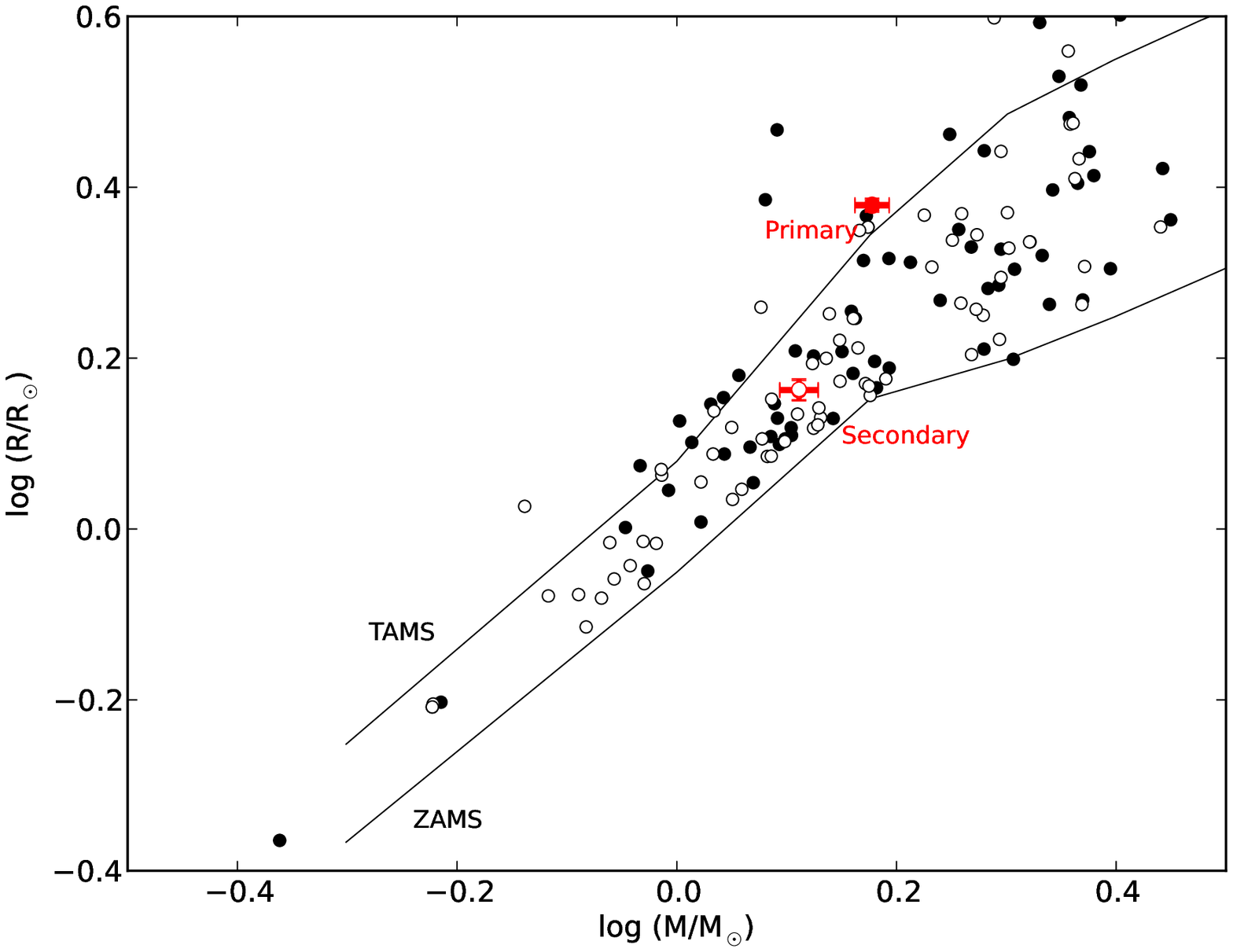}
  \label{fig7b}
\end{minipage}
\caption{Mass-Luminosity (left panel) and Mass-Radius (right panel) diagrams. 
ZAMS and TAMS computed from the theoretical models by \citet{mowlavi2012} are 
plotted by solid lines.  Positions of components in selected detached systems 
from \citet{torres2010} are also shown. Symbols mean the same as in Fig. \ref{fig6}.} 
\label{fig7}
\end{figure}

The positions of \astrobj{V482~Per}'s components in all evolution diagrams in 
Figs. \ref{fig6} \& \ref{fig7} indicate the primary to be somewhat evolved 
(just above the TAMS) and more luminous than expected for its mass. 
This statement is true for solutions obtained for both spectral types assumed. 
Based on the position of the secondary (see Figs. \ref{fig6} and \ref{fig7}), it is
evolving on the Main Sequence, about half way  between ZAMS and TAMS.
It seems that a better agreement with the evolutionary models was obtained
for the hotter, F2 type primary model. Within the F5 solution the secondary is
significantly underlumious (see Figs. \ref{fig6} and \ref{fig7}). 


Our analysis of the system's O-C diagram provides a strong evidence for the existence
of a third body, proposed in earlier works of \citet{wolfetal04}, 
and \citet{ogloza12}. 
Knowing the absolute parameters of the system's components, and assuming a coplanar orbit with the binary, we found 
the companion to have a very close lower mass limit (2.14 M$_{\odot}$) to their 
estimates based on guessed absolute parameters of the system.
The orbital period ($\sim$16 years), eccentricity ($\sim$0.83) and the 
size of the gravitationally bound component's orbit are also very similar in 
both of these studies and ours. We confirm their findings based on 
the absolute parameters determined from combined spectroscopy and photometry
analysis.
\begin{table*}
\small
\begin{center}
\caption{Absolute Parameters}
\label{table4}
\medskip
\begin{tabular}{lcccc}
\hline
\hline
 & \multicolumn{2}{c}{F2 model} & \multicolumn{2}{c}{F5 model} \\
Parameter & Value & Error & Value & Error \\
\hline
$a [R_{\odot}]$ & 10.76 & 0.16 & 10.76 & 0.16 \\
$M_{1} [M_{\odot}]$ & 1.51 & 0.05 & 1.51 & 0.05 \\
$M_{2} [M_{\odot}]$ & 1.29 & 0.05 & 1.29 & 0.05 \\
$R_{1} [R_{\odot}]$ & 2.39 & 0.04 & 2.40 & 0.04 \\
$R_{2} [R_{\odot}]$ & 1.45 & 0.04 & 1.46 & 0.04 \\
$L_{1} [L_{\odot}]$ & 10.15 & 0.69 & 8.85 & 0.62 \\
$L_{2} [L_{\odot}]$ & 3.01 & 0.17 & 2.65 & 0.15 \\
\hline
\end{tabular}
\end{center}
\begin{flushleft}
\noindent {\bf Note:} $a$ - orbital semi-major axis in units of  solar radius, 
$M_{1,2}$ - masses, $R_{1,2}$ - radii, $L_{1,2}$ - luminosities of the primary
and secondary, respectively.
\end{flushleft}
\end{table*}
The 2.14 M$_{\odot}$ tertiary companion would be an A1 type star with about
9100~K temperature and a radius of 2.02 R$_{\odot}$. Its luminosity
would be about 25.5~L$_{\odot}$, twice exceeding that of the binary
system. However, we found a very small contribution reaching about 7\%
in the I filter and decreasing towards shorter wavelengths. 
From a  crude estimate it follows that the companions should have about
1~L$_{\odot}$ in the near infrared and smaller in shorter wavelengths.
Unfortunately, we haven't seen a trace of a third body in the spectra which might be due to our resolution limit.
As an explanation, a binary system consising of a dwarf and a compact object
would agree with our findings. 
Further radial velocity observations are needed to verify this hypothesis.

\section*{Acknowledgments}

The authors acknowledge the use of the Simbad database, operated at the CDS, Strasbourg, 
France, and of NASA's Astrophysics Data System Bibliographic Services. We would 
like to thank all the staff working for Dominion Astrophysical Observatory and 
Gerostathopoulion Observatory of the University of Athens. SZ gratefully acknowledges the support by the NCN grant No. 2012/07/B/ST9/04432.



\newpage
\appendix
\section{}
\label{appendix}
\setcounter{table}{0} \renewcommand{\thetable}{A.\arabic{table}}
\scriptsize
\begin{longtable}{cccc|cccc}
\caption{Eclipse Timings for \astrobj{V482~Per}}
\label{apptable}\\
\hline
\hline
HJD-2400000 & Type & Method & Ref & HJD-2400000 & Type & Method & Ref\\  
\hline
\endfirsthead
\multicolumn{8}{c}%
        {{Table \thetable\ Continued from the previous page}} \\
\hline
\hline
\endhead
47565.3737 & II & pe & 1 & 53318.9165 & I & ccd & 18\\
47823.5048 & I & pe & 1 & 53656.5685 & I & ccd & 19\\
47840.6360 & I & pe & 1 & 53668.8015 & I & ccd & 18\\
47850.4210 & I & pe & 1 & 53685.9293 & I & ccd & 18\\
47943.4012 & I & pe & 1 & 53739.7567 & I & ccd & 20\\
48222.3268 & I & pe & 1 & 53744.6506 & I & ccd & 20\\
48222.3276 & I & pe & 2 & 53766.6715 & I & ccd & 20\\
48299.4004 & II & pe & 2 & 53793.5857 & I & ccd & 20\\
48606.4669 & I & pe & 3 & 54055.3865 & I & ccd & 21\\
48650.5058 & I & pe & 3 & 54057.8341 & I & ccd & 20\\
49060.3247 & II & pe & 4 & 54073.7380 & II & ccd & 20\\
49372.2657 & I & pe & 5 & 54193.6316 & II & ccd & 22\\
49625.5031 & II & pe & 6 & 54400.3747 & I & pe & 23\\
49761.3046 & I & pe & 7 & 54402.8237 & I & ccd & 22\\
50047.5745 & I & ccd & 7 & 54408.9423 & II & ccd & 22\\ 
50079.3873 & I & ccd & 7 & 54433.4101 & II & pe & 23\\ 
50106.2985 & I & ccd & 8 & 54515.3736 & I & ccd & 24\\
50380.3383 & I & pe & 9 & 54764.9392 & I & ccd & 22\\
51185.3277 & I & pe & 10 & 54764.9399 & I & ccd & 22\\
51373.7390 & I & ccd & 11 & 54774.7265 & I & ccd & 22\\
52250.9000 & II & ccd & 12 & 54786.9621 & I & ccd & 22\\ 
52266.8056 & I & ccd & 12 & 54801.6414 & I & ccd & 22\\
52276.5957 & I & ccd & 12 & 54812.6523 & II & ccd & 22\\ 
52287.6027 & II & ccd & 12 & 54816.3222 & I & ccd & 25\\ 
52288.8255 & I & ccd & 12 & 54840.7874 & I & ccd & 22\\
52589.7781 & I & ccd & 12 & 54840.7884 & I & ccd & 22\\
52644.8308 & II & ccd & 13 & 54845.6802 & I & ccd & 22\\
52724.3527 & I & ccd & 14  & 54845.6826 & I & ccd & 26\\
52907.8584 & I & ccd & 13 & 54867.7018 & I & ccd & 22\\ 
52949.4538 & I & ccd & 15 & 54867.7024 & I & ccd & 22\\
52949.4565 & I & ccd & 16 & 55158.8543 & I & ccd & 27\\
53032.6433 & I & pe & 17 & 55169.8635 & II & ccd & 27\\
53317.6900 & II & ccd & 18 & 55185.7682 & I & ccd & 27\\ 
55201.6755 & II & ccd & 27 & 55563.7765 & II & ccd & 27\\
55245.7097 & II & ccd & 27 & 55573.5628 & II & ccd & 27 \\
55432.8751 & I & ccd & 27 & 55579.6804 & I & ccd & 29\\
55443.8859 & II & ccd & 27 & 55590.6906 & II & ccd & 27\\
55459.7893 & I & ccd & 27 & 55599.2520 & I & ccd & 30\\ 
55476.9162 & I & ccd & 27 & 55847.6067 & II & ccd & 31\\ 
55476.9168 & I & ccd & 28 & 55848.8258 & I & ccd & 32\\ 
55481.8093 & I & ccd & 27 & 55848.8265 & I & ccd & 32\\ 
55497.7146 & II & ccd & 27 & 55852.4969 & II & ccd & 31\\ 
55498.9346 & I & ccd & 27 & 55865.9522 & I & ccd & 32\\ 
55498.9364 & I & ccd & 27 & 55865.9527 & I & ccd & 32\\ 
55503.8308 & I & ccd & 27 & 55868.3970 & I & ccd & 31\\ 
55508.7236 & I & ccd & 27 & 55875.7404 & I & ccd & 32\\ 
55509.9476 & II & ccd & 27 & 55940.5796 & II & ccd & 32\\ 
55514.8421 & II & ccd & 27 & 55946.6960 & I & ccd & 32\\ 
55519.7355 & II & ccd & 27 & 56221.9605 & II & ccd & 33\\ 
55536.8653 & II & ccd & 27 & 56232.9677 & I & ccd & 34\\ 
55557.6609 & I & ccd & 27 & 56324.7231 & II & ccd & 33\\ 
55563.7761 & II & ccd & 27 & & & & \\ 
\hline
\end{longtable}
\begin{flushleft}
\noindent {\bf References:} $^{1}$\citet{agererlicht91}, $^{2}$\citet{hubscher91}, $^{3}$\citet{hubscher92}, $^{4}$\citet{hubscher93}, $^{5}$\citet{hubscher94}, $^{6}$\citet{agererhubscher95}, $^{7}$\citet{agererhubscher96}, $^{8}$\citet{agererhubscher97}, $^{9}$\citet{agererhubscher98}, $^{10}$\citet{agerer99}, $^{11}$\citet{paschke99}, $^{12}$\citet{lacy2002}, $^{13}$\citet{lacy2003}, $^{14}$\citet{agererhubscher03}, $^{15}$\citet{kotkovawolf06}, $^{16}$\citet{zejda2004}, $^{17}$\citet{lacy2004}, $^{18}$\citet{lacy2006}, $^{19}$\citet{brat2007}, $^{20}$\citet{lacy2007}, $^{21}$\citet{hubscherwalter07}, $^{22}$\citet{lacy2009}, $^{23}$\citet{yilmaz2009}, $^{24}$\citet{hubscher2009a}, $^{25}$\citet{hubscher2009b}, $^{26}$\citet{diethelm2009}, $^{27}$\citet{lacy2011}, $^{28}$\citet{diethelm2011a}, $^{29}$\citet{diethelm2011b}, $^{30}$\citet{hubscher2012}, $^{31}$\citet{liakos2011}, $^{32}$\citet{lacy2012}, $^{33}$\citet{lacy2013}, $^{34}$\citet{diethelm2013}
\end{flushleft}

\begin{thebibliography}{ }
\bibitem[\protect\citeauthoryear{Agerer et al.}{1999}]{agerer99} Agerer, F., Dahm, M., H\"ubscher, J., 1999, IBVS, 4712
\bibitem[\protect\citeauthoryear{Agerer \& H\"ubscher}{1995}]{agererhubscher95} Agerer, F., H\"ubscher, J., 1995, IBVS, 4222
\bibitem[\protect\citeauthoryear{Agerer \& H\"ubscher}{1996}]{agererhubscher96} Agerer, F., H\"ubscher, J., 1996, IBVS, 4383
\bibitem[\protect\citeauthoryear{Agerer \& H\"ubscher}{1997}]{agererhubscher97} Agerer, F., H\"ubscher, J., 1997, IBVS, 4472
\bibitem[\protect\citeauthoryear{Agerer \& H\"ubscher}{1998}]{agererhubscher98} Agerer, F., H\"ubscher, J., 1998, IBVS, 4562
\bibitem[\protect\citeauthoryear{Agerer \& H\"ubscher}{2003}]{agererhubscher03} Agerer, F., H\"ubscher, J., 2003, IBVS, 5484
\bibitem[\protect\citeauthoryear{Agerer \& Lichtenknecker}{1991}]{agererlicht91} Agerer, F., Lichtenknecker, D., 1991, IBVS, 3554
\bibitem[\protect\citeauthoryear{Br\'at et al.}{2007}]{brat2007} Br\'at, L., Zejda, M., Svoboda, P., 2007, OEJV, 74
\bibitem[\protect\citeauthoryear{Claret \& Bloemen}{2011}]{Claret2011} Claret, A., Bloemen, S., 2011, A\&A, 529, 75.
\bibitem[\protect\citeauthoryear{Claret et al.}{2012}]{Claret2012} Claret, A., Hauschildt, P. H., Witte, S., 2012, A\&A, 546, 14
\bibitem[\protect\citeauthoryear{Claret et al.}{2013}]{Claret2013} Claret, A., Hauschildt, P. H., Witte, S., 2013, A\&A, 552, 16
\bibitem[\protect\citeauthoryear{Diethelm}{2009}]{diethelm2009} Diethelm, R.,2009, IBVS, 5894
\bibitem[\protect\citeauthoryear{Diethelm}{2011a}]{diethelm2011a} Diethelm, R., 2011a, IBVS, 5960
\bibitem[\protect\citeauthoryear{Diethelm}{2011b}]{diethelm2011b} Diethelm, R., 2011b, IBVS, 5992
\bibitem[\protect\citeauthoryear{Diethelm}{2013}]{diethelm2013} Diethelm, R., 2013, IBVS, 6042
\bibitem[\protect\citeauthoryear{Ekstr\"{o}m et al.}{2012}]{ekstrom2012} Ekstr\"{o}m, S., Georgy, C., Eggenberger, P., Meynet, G., Mowlavi, N., Wyttenbach, A. , Granada, A., Decressin, T., Hirschi, R., Frischknecht, U.,, Charbonnel, C., Maeder A., 2012, A\&A, 537, A146.
\bibitem[\protect\citeauthoryear{Harmanec}{1988}]{Harmanec} Harmanec, P., 1988, BAICz, 39, 329.
\bibitem[\protect\citeauthoryear{Harvig \& Leis}{1981}]{harvig81} Harvig, V., Leis L., 1981, PTarO, 48, 172.
\bibitem[\protect\citeauthoryear{Hoffmeister}{1966}]{hoff66} Hoffmeister, C., 1966, AN, 289, 1.
\bibitem[\protect\citeauthoryear{Hroch}{1998}]{hroch98} Hroch, F., 1998, Proceedings of the 29th Conference on Variable Star Research, 30.
\bibitem[\protect\citeauthoryear{H\"ubscher et al.}{1991}]{hubscher91} H\"ubscher, J., Agerer, F., R\"uckersdorf, E.W., 1991, BAVSM, 59
\bibitem[\protect\citeauthoryear{H\"ubscher et al.}{1992}]{hubscher92} H\"ubscher, J., Agerer, F., Wunder, E., 1992, BAVSM, 60
\bibitem[\protect\citeauthoryear{H\"ubscher et al.}{1993}]{hubscher93} H\"ubscher, J., Agerer, F., Wunder, E., 1993, BAVSM, 62
\bibitem[\protect\citeauthoryear{H\"ubscher et al.}{1994}]{hubscher94} H\"ubscher, J., Agerer, F., Frank, P., Wunder, E., 1994, BAVSM, 68
\bibitem[\protect\citeauthoryear{H\"ubscher et al.}{2009a}]{hubscher2009a} H\"ubscher, J., Steinbach, H.-M., Walter, F., 2009a, IBVS, 5874
\bibitem[\protect\citeauthoryear{H\"ubscher et al.}{2009b}]{hubscher2009b} H\"ubscher, J., Steinbach, H.-M., Walter, F., 2009b, IBVS, 5889
\bibitem[\protect\citeauthoryear{H\"ubscher et al.}{2012}]{hubscher2012} H\"ubscher, J., Lehmann, P.~B., Walter, F., 2012, IBVS, 6010
\bibitem[\protect\citeauthoryear{H\"ubscher \& Walter}{2007}]{hubscherwalter07} H\"ubscher, J., Walter, F., 2007, IBVS, 5761
\bibitem[\protect\citeauthoryear{Irwin}{1952}]{irwin52} Irwin, J. B., 1952, ApJ, 116, 211
\bibitem[\protect\citeauthoryear{Irwin}{1959}]{irwin59} Irwin, J. B., 1959, AJ, 64, 149
\bibitem[\protect\citeauthoryear{Kallrath et al.}{1998}]{kallrath98} Kallrath, J., Milone, E. F., Terrell, D. Y., Andrew T., 1998, ApJ, 508, 308
\bibitem[\protect\citeauthoryear{Lacy}{2002}]{lacy2002} Lacy, C.H.S., 2002, IBVS, 5357
\bibitem[\protect\citeauthoryear{Lacy}{2003}]{lacy2003} Lacy, C.H.S., 2003, IBVS, 5487
\bibitem[\protect\citeauthoryear{Lacy}{2004}]{lacy2004} Lacy, C.H.S., 2004, IBVS, 5577
\bibitem[\protect\citeauthoryear{Lacy}{2006}]{lacy2006} Lacy, C.H.S., 2006, IBVS, 5670
\bibitem[\protect\citeauthoryear{Lacy}{2007}]{lacy2007} Lacy, C.H.S., 2007, IBVS, 5764
\bibitem[\protect\citeauthoryear{Lacy}{2009}]{lacy2009} Lacy, C.H.S., 2009, IBVS, 5910
\bibitem[\protect\citeauthoryear{Lacy}{2011}]{lacy2011} Lacy, C.H.S., 2011, IBVS, 5972
\bibitem[\protect\citeauthoryear{Lacy}{2012}]{lacy2012} Lacy, C.H.S., 2012, IBVS, 6014
\bibitem[\protect\citeauthoryear{Lacy}{2013}]{lacy2013} Lacy, C.H.S., 2013, IBVS, 6046
\bibitem[\protect\citeauthoryear{Kotkova \& Wolf}{2006}]{kotkovawolf06} Kotkova, L., Wolf, M., 2006, IBVS, 5676
\bibitem[\protect\citeauthoryear{Liakos \& Niarchos}{2011}]{liakos2011} Liakos, A., Niarchos, P., 2011, IBVS, 6005
\bibitem[\protect\citeauthoryear{Motl}{2004}]{motl04} Motl, D., 2004, C-MUNIPACK, http://c-munipack.sourceforge.net/.
\bibitem[\protect\citeauthoryear{Mowlavi et al.}{2012}]{mowlavi2012} Mowlavi, N., Eggenberger, P., Meynet, G., Ekström, S., Georgy, C., Maeder, A., Charbonnel, C., Eyer, L., 2012,  A\&A, 541, 41.
\bibitem[\protect\citeauthoryear{Nelson et al.}{2006}]{nelson2006} Nelson, R.H., Terrell, D., Gross, J., 2006. IBVS 5715, 1.
\bibitem[\protect\citeauthoryear{Nelson}{2010}]{nelson2010a} Nelson, R.H., 2010a, Software by Bob Nelson. (http://members.shaw.ca/bob.nelson/software1.htm).
\bibitem[\protect\citeauthoryear{Nelson}{2010}]{nelson2010b} Nelson, R.H., 2010b, 'Spectroscopy for Eclipsing Binary Analysis' in The Alt-Az Initiative, Telescope Mirror \& Instrument Developments (Collins Foundation Press, Santa Margarita, CA). R.M. Genet, J.M. Johnson and V. Wallen (eds).
\bibitem[\protect\citeauthoryear{Ogloza et al.}{2012}]{ogloza12} Ogłoza, W., Kreiner, J.~M., Stachowski, G., Winiarski, M., Zakrzewski, B., Dogru, S., Alicavus, F., Demircan, O., Erdem, A., 2012, IAUS, 282, 850.
\bibitem[\protect\citeauthoryear{Popper}{1996}]{popper96} Popper, D.~M., 1996, ApJS, 106, 133.
\bibitem[\protect\citeauthoryear{Paschke}{1999}]{paschke99} Paschke, A., 1999, O-C Gateway, http://var.astro.cz/ocgate/
\bibitem[\protect\citeauthoryear{Rucinski}{2004}]{rucinski04} Rucinski, S.M., 2004. IAUS 215, 17.
\bibitem[\protect\citeauthoryear{Torres et al.}{2010}]{torres2010} Torres, G., Andersen, J., Gim\'enez, A., 2011, A\&ARv, 18, 67
\bibitem[\protect\citeauthoryear{van Hamme}{1993}]{vanHamme93} van Hamme, W., 1993, AJ, 106, 296
\bibitem[\protect\citeauthoryear{Wilson \& Devinney}{1971}]{wilsondev71} Wilson, R.~E., Devinney, E.~J, 1971, ApJ, 166, 605.
\bibitem[\protect\citeauthoryear{Wilson}{1990}]{wilson90} Wilson, R.~E., 1990, ApJ, 356, 613.
\bibitem[\protect\citeauthoryear{Wolf et al.}{2004}]{wolfetal04} Wolf, M., Mayer, P., Zasche, P., Sarounova, L., Zejda, M., 2004, ASPC, 318, 255.
\bibitem[\protect\citeauthoryear{Y{\i}lmaz et al.}{2009}]{yilmaz2009} Y{\i}lmaz, M., Ba\c{s}t\"urk, \"O., Alan, N., and 13 co-authors, 2009, IBVS, 5887
\bibitem[\protect\citeauthoryear{Zejda}{2004}]{zejda2004} Zejda, M., 2004, IBVS, 5583
\bibitem[\protect\citeauthoryear{Zola et al.}{2014}]{zolaetal2014} Zola, S., \c{S}enavc{\i}, H.~V., Liakos, A., Nelson, R.~H., Zakrzewski, B., 2014, MNRAS, 437, 3718.
\end{thebibliography}
\end{document}